\documentstyle[preprint,aps,epsfig,psfig]{revtex}

\begin{document}

\title{Further Analysis of Di-gluon Fusion Mechanism \\
for the decays of $B\rightarrow K \eta^{\prime}$}
\vspace{2cm}

\author{
Dongsheng Du${}^{1,2}$, Deshan Yang${}^{2}$  and Guohuai Zhu${}^{2}$
\footnote{Email: duds@hptc5.ihep.ac.cn, yangds@hptc5.ihep.ac.cn,
zhugh@hptc5.ihep.ac.cn} }
\address{${}^1$ CCAST (World Laboratory), P.O.Box 8730, Beijing
100080, China\\
${}^2$ Institute of High Energy Physics, Chinese Academy of Sciences,
 P.O.Box 918(4), Beijing 100039, China
 \footnote{Mailing address}}

\maketitle
\begin{abstract}
Di-gluon fusion mechanism might
account for the large branching ratios of $B\rightarrow K \eta^{\prime}$.
But because we know little about the effective $\eta^{\prime}gg$ vertex,
there are large uncertainties in perturbative QCD estimations. In this paper,
we try several kinds of $\eta^{\prime}gg$ form factors and compare the
numerical results with the experiment. We find that, though we know little
about $\eta^{\prime}gg$ form factor, if di-gluon fusion mechanism is
important in $B\rightarrow K \eta^{\prime}$,
the branching ratios of the decays 
$B\rightarrow K^{\ast} \eta^{\prime}$ 
would be around $10^{-5}$ which can be tested by future experiments.
\end{abstract}

\vspace{1.5cm}

{\bf PACS numbers 13.25.Hw 13.20.He}

\newpage

\tighten
\narrowtext

\section{Introduction}
Recently, CLEO collaboration have
improved their previous measurements of $B\rightarrow K \eta^{\prime}$
\cite{CLEO}:
\begin{eqnarray}
{\cal{B}}(B^{+}\rightarrow K^{+} \eta^{\prime})=(80^{+10}_{-9}\pm 8)
\times 10^{-6}
\nonumber\\
{\cal{B}}(B^{0}\rightarrow K^{0} \eta^{\prime})=(88^{+18}_{-16}\pm 9)
\times 10^{-6}.
\end{eqnarray}
These branching ratios are much larger than the estimations under the
standard theorectical frame which is based on the effective Hamiltonian and
general factorization approximation. Now
it is commonly believed that these large branching ratios are due to the
special properties of $\eta^{\prime}$, and several new mechanisms have been
proposed to enhance the decay rates of $B \rightarrow \eta^{\prime} K$ and
$B \rightarrow \eta^{\prime} X_s$. In the following we only discuss the
standard model(SM) mechanisms because we think that the contribution of the
SM should be carefully examined first.

Halperin and Zhitnisky \cite{Zhitnisky} proposed an interesting mechanism:
$\eta^{\prime}$ can be directly produced through
$b \rightarrow \bar c cs \rightarrow \eta^{\prime}s$ if
$\langle \eta^{\prime}|\bar c \gamma_{\mu}\gamma_5 c|0 \rangle = -i
f^c_{\eta^{\prime}} P^{\mu}_{\eta^{\prime}} \not= 0$.
But if this mechanism dominates, one easily find that
\begin{eqnarray}
\frac{{\cal{B}}(B^0 \rightarrow \eta^{\prime} K^{\ast 0})}
{{\cal{B}}(B^0 \rightarrow \eta^{\prime} K^{0})} &=&
\frac{|\langle \eta^{\prime}|(\bar c c)_{V-A}|0\rangle \langle K^{\ast 0}|
(\bar s b)_{V-A}|B^0\rangle|^2}
{|\langle \eta^{\prime}|(\bar c c)_{V-A}|0\rangle \langle K^{0}|
(\bar s b)_{V-A}|B^0\rangle|^2} \nonumber \\
& = & \frac{|2 f^c_{\eta^{\prime}} M_{K^{\ast}} 
A_0^{BK^{\ast}}(M^2_{\eta^{\prime}}) (\epsilon_{K^{\ast}} \cdot P_B)|^2}
{|i f^c_{\eta^{\prime}} (M_B^2 - M^2_K) F_0^{BK}(M^2_{\eta^{\prime}})|^2}
\simeq \left ( \frac{A_0^{BK^{\ast}}(M^2_{\eta^{\prime}})}
{F_0^{BK}(M^2_{\eta^{\prime}})} \right )^2 \simeq 0.9,
\end{eqnarray}
which is in contradiction with the stringent upper limit of
$B^0 \rightarrow \eta^{\prime} K^{\ast 0}$ given by CLEO \cite{CLEO}:
\begin{eqnarray}
{\cal{B}}(B^0 \rightarrow \eta^{\prime} K^{\ast 0}) < 2.0 \times 10^{-5}.
\end{eqnarray}
So this mechanism is unlikely to be dominant.
Because the effective vertex of $b \rightarrow sgg$ in SM is very small,
$b \rightarrow sgg \rightarrow s \eta^{\prime}$ \cite{He} is also impossible
to account for the large $B \rightarrow \eta^{\prime} K$ branching ratios.
The authors of Ref \cite{AS} proposed that $b \rightarrow s g^{\ast}$ and
$g^{\ast} \rightarrow g \eta^{\prime}$ via QCD anomaly can account for the
large semi-inclusive branching ratios of $B \rightarrow \eta^{\prime} X_s$.
CLEO \cite{CLEO1} have measured the $\eta^{\prime}$ momentum spectrum in
$B \rightarrow \eta^{\prime} X_s$ and this measurement favors the mechanism
of Ref\cite{AS}. But because it has an extra gluon in the final state, 
unless the gluon is soft and absorbed into the $\eta^{\prime}$ wave 
functions \cite{chenghy}, it 
seems difficult to realize its contribution to two-body exclusive decay
$B \rightarrow \eta^{\prime} K$.

Motivated by the idea of Ref \cite{AS,HT}, the authors of Ref \cite{yangyd}
proposed di-gluon fusion mechanism, which is depicted in Fig.1.
Because both of the gluons are hard, it seems reasonable to give a
perturbative QCD estimation. But because of our ignorance about the form
factor of $\eta^{\prime}gg$ vertex, there are large uncertainties
in calculations. So in the following we would reanalyze di-gluon fusion
mechanism in detail using several kinds of 
$\langle gg | \eta^{\prime} \rangle$ form 
factors and compare our numerical results with experimental data.

\section{Analysis of di-gluon fusion mechanism}
We first give a brief introduction on how to estimate contributions of 
di-gluon fusion mechanism in perturbative QCD.

$g-g-\eta^{\prime}$ vertex can be parameterized as \cite{Close}:
\begin{equation}
\langle g^a g^b \vert \eta^{\prime} \rangle = 4 \delta^{ab} g_s^2
A_{\eta^{\prime}} F(k_1^2, k_2^2) \epsilon_{\mu \nu \rho \sigma} k_1^{\mu} 
k_2^{\nu} \epsilon_1^{\rho} \epsilon_2^{\sigma},
\end{equation}
where $A_{\eta^{\prime}}$ is a constant and can be extracted from
$J/\Psi \rightarrow \gamma \eta^{\prime}$ or
$\eta^{\prime}\rightarrow \gamma \gamma$,
$g_s$ is QCD coupling constant, 
$F(k_1^2,k_2^2)$ is $\langle gg | \eta^{\prime} \rangle$ form factor and 
$F(0,0)=1$.
Neglecting the transverse momentum of quarks and taking the wave functions
for $B$ and $K$ meson as
\begin{eqnarray}
\Psi_B=\frac{1}{\sqrt {2}} \frac{I_C}{\sqrt{3}} \phi_B(x)
(\slash{\hskip -2mm}p+m_B)\gamma_5, &~~~&
\Psi_K=\frac{1}{\sqrt {2}} \frac{I_C}{\sqrt{3}} \phi_K(y)
\slash{\hskip -2mm}q \gamma_5,
\end{eqnarray}
where $I_C$ is an identity in color space, and the integration of the 
distribution amplitude is related to the meson decay constant,
\begin{equation}
\int \phi_B(x)dx=\frac{f_B} {2\sqrt{6}},~~~~
\int \phi_K(y)dy=\frac{f_K}{2\sqrt{6}}.
\end{equation}
In this paper, we take the distribution amplitudes as\cite{wf}:
\begin{equation}
\phi_B(x)=\frac{1}{2\sqrt{6}} \delta(x-\epsilon_B),~~~~~~~~
\phi_K(y)=\sqrt{\frac{3}{2}} y(1-y).
\end{equation}
Then we can write the amplitude of Fig.1 as 
\begin{equation}
{\cal{M}}=\int dxdy \phi_B(x)
\frac{Tr[\gamma_5 \slash{\hskip -2mm}q \Gamma_{\mu}
(\slash{\hskip -2mm}p+m_B)\gamma_5 \gamma_{\nu}]
4 \epsilon^{\mu \nu \alpha \beta} k_{1\alpha} k_{2\beta}
A_{\eta^{\prime}} F(k_1^2,k_2^2) C_F g_s^3}
{\sqrt{2} \sqrt{2} k_1^2 k_2^2}
\phi_K(y),
\end{equation}
where $\Gamma_{\mu}$ is the effective $b\rightarrow sg$ vertex\cite{Hou}
\begin{equation}
\Gamma_{\mu}^a=\frac{G_F}{\sqrt{2}} \frac{g_s}{4 \pi^2} 
V_{is}^*V_{ib} \bar{s} t^a\left\{F_1^i(k_1^2\gamma_{\mu}-k_{1\mu} 
\slash{\hskip -2mm}k_1)\frac{1-\gamma_5}{2}-F_2^i i \sigma_{\mu \nu}
k_1^{\nu} m_b \frac{1+\gamma_5}{2} \right\}b,
\end{equation}
and the color factor
$C_F=\frac{1}{\sqrt{3}} \frac{1}{\sqrt{3}} Tr[t^a t^b]\delta^{ab}=\frac{4}{3}$.
Finally we obtain 
\begin{eqnarray}
{\cal{M}}&=&\frac{G_F}{\sqrt{2}} \alpha_s^2 A_{\eta^{\prime}} 
C_F 32 V_{is}^* V_{ib}
\int dxdy \phi_B(x) \phi_K(y) F(k_1^2,k_2^2) \nonumber \\
&~&\times 
\frac{F_1^i k_1^2[(p\cdot k_1)(q\cdot k_2)-(p\cdot k_2)(q \cdot k_1)]
+F_2^i m_B m_b[(q\cdot k_2) k_1^2-(q\cdot k_1)(k_1\cdot k_2)]}
{k_1^2 k_2^2} .
\end{eqnarray}
In the above integral, there are two singularities from the gluon
propagators and sometimes there would be another singularity from the
$\langle gg | \eta^{\prime} \rangle $ form factor (For example,
$F(k_1^2,k_2^2)=\frac{m^2_{\eta^{\prime}}}{2(k_1\cdot k_2)}$).
We add a small pure imaginary number $i\epsilon$ ($\epsilon > 0$) 
to the denominator to get a convergent integral.

To evaluate the numerical result of Eq.(10), we take the form factors
$F_1$ and $F_2$ according to ref\cite{AS,He1} as
\begin{eqnarray}
F_1=\frac{4\pi}{\alpha_s}(C_4+C_6),~~~~~~F_2=-2 C_8.
\end{eqnarray}
Where $C_i$ are wilson coefficients at the NLL level.(Because QCD
corrections maybe large, we do not take \cite{HT,Hou} $F_1 \simeq -5$ and
$F_2 \simeq 0.2$ which are derived from the SM without QCD corrections.)
It is always subtle to choose the scale of $\alpha_s$, because the average 
momenta squared of the gluons are
\begin{eqnarray}
\langle k_1^2 \rangle \simeq 12~GeV^2 ~~~~~~
\langle k_2^2 \rangle \simeq 1~GeV^2 
\end{eqnarray}
we prefer to choose 
$\alpha_s=\sqrt{\alpha_s(k_1^2)\alpha_s(k_2^2)}=0.28$ 
though we also give the branching ratios when 
$\alpha_s=\alpha_s(k_1^2)=0.21$ or $\alpha_s=\alpha_s(k_2^2)=0.38$
in Tab.1 and Tab.2.

As to the $\langle gg | \eta^{\prime} \rangle$ form
factor $F(k_1^2,k_2^2)$, because of our complete ignorance there are large
uncertainties in the numerical
evaluation of Eq.(10). In the following we try several kinds of form factors.

In Ref\cite{AS,HT}, the authors have assumed that $q^2$-dependence of
the form factor is weak and as an approximation they take
form factor as a constant to explain large semi-inclusive decay
$B\rightarrow X_s \eta^{\prime}$. The difference between Ref\cite{AS}
and Ref\cite{HT} is that the running of $\alpha_s(\mu)$ with the scale is 
considered in \cite{HT} but not in \cite{AS}. 
We also use their ansatz to estimate
exclusive decays $B\rightarrow K \eta^{\prime}$ and
$B\rightarrow K^{\ast} \eta^{\prime}$. Using Eq.(10), we give 
numerical results in Tab.1. From the Table it seems that the constant
form factor can account for the large branching ratios of
$B\rightarrow K \eta^{\prime}$, but unfortunately we would get much
larger branching ratios of $B\rightarrow K^{\ast} \eta^{\prime}$, which
is strongly in contradiction with the CLEO's measurements\cite{CLEO}
(see, for instance, the case of $\epsilon_B=0.06$ in Tab.1).

From Eq.(10), we notice that the amplitude must be integrated over a
wide range of $k_1^2$ and $k_2^2$, therefore the effects of
$k_1^2$, $k_2^2$
dependence of $F(k_1^2,k_2^2)$ must be taken into account.

The authors of Ref\cite{yangyd} choose the form factor as
$F(k_1^2,k_2^2)=\frac{m^2_{\eta^{\prime}}}{2(k_1\cdot k_2)}$ because
such a form factor works well in $J/\Psi\rightarrow \gamma \eta^{\prime}$
\cite{kuhn}. We have followed their calculations and it seems that their
numerical results are overestimated than ours(in Tab.2) by a factor about
three. From Tab.2, we can see that our estimations of
${\cal{B}}(B\rightarrow \eta^{\prime} K)$ are about $10^{-5}$, but when
considering that the standard theoretical frame (based on effective
Hamiltonian and general factorization approximation) can give\cite{chenghy}
${\cal{B}}(B\rightarrow \eta^{\prime} K)\simeq 3.6\times 10^{-5}$, it is
still possible to account for the experimental data only if the contributions
from di-gluon fusion mechanism and the standard theoretical frame 
constructively interfere. If this is true, then because the contributions
to ${\cal{B}}(B\rightarrow \eta^{\prime} K^{\ast})$ from the standard 
theoretical frame are negligible, the di-gluon
fusion mechanism would be dominant in
$B\rightarrow \eta^{\prime} K^{\ast}$, {\it{i.e.}}, 
${\cal{B}}(B\rightarrow \eta^{\prime} K^{\ast}) \sim 10^{-5}$ which can be
tested by future measurements of CLEO or B factories.

Kagan and Petrov \cite{kagan} proposed a model of the $g-g-\eta^{\prime}$
vertex in which a pseudoscalar current is coupled to two gluons through
quark loops. Their perturbative calculations yield a form factor:
\begin{eqnarray}
F(k_1^2,k_2^2)&\propto& \sum \limits_{f=u,d,s} a_f m_f\nonumber \\
&& \times \int \limits_{0}^{1} dx \int \limits_{0}^{1-x} 
\frac{dy} {m_f^2-(1-x-y)(xk_1^2+yk_2^2)-x y m_{\eta^{\prime}}^2-i\epsilon},
\end{eqnarray}
where we take $a_u=a_d=1$, $a_s=2$ and normalize $F(k_1^2,k_2^2)$ with the 
normal condition $F(0,0)=1$. We use this form factor in Eq.(10) with
$\epsilon_B=0.06$ and get 
\begin{eqnarray}
{\cal{B}}(B \rightarrow \eta^{\prime} K)=1.28 \times 10^{-7}.
\end{eqnarray}
which is too small to account for the experiments.

In Ref \cite{ong}, the authors calculate the transition form factor in 
$\pi^{0}$ coupling to $\gamma^{*} \gamma^{*}$ in the frame of a perturbative
QCD based on the modified factorization formula.
They find that numerically their results are extremely
similar to that obtained by applying the interpolation procedure in the manner
of Brodsky and Lepage in the case of two off-shell photons:
\begin{equation}
F_{\pi^{0}\gamma^{*}\gamma^{*}}(k_1^2,k_2^2)
\propto (1-\frac{X^2}{\Lambda_{\pi}^2})^{-1},
\end{equation}
with 
\begin{equation}
X^2=\frac{(k_1^2-k_2^2)^3}{k_1^4-k_2^4-2 k_1^2 k_2^2 \ln(k_1^2/k_2^2)}
\end{equation}
and $\Lambda_{\pi}=2 \pi f_{\pi}=0.83GeV$.
We assume that the structure of the transition of
$\eta^{\prime}\rightarrow g^{*} g^{*}$ is similar to that of 
$\pi^0 \rightarrow \gamma^{*} \gamma^{*}$:
\begin{equation}
F(k_1^2,k_2^2)=(1-\frac {X^2} {\Lambda_{\eta^{\prime}}^2})^{-1},
\end{equation}
where we approximate $\Lambda_{\eta^{\prime}}=\Lambda_{\pi}$.
By using this form factor, we calculate the branching ratios of
$B\rightarrow K \eta^{\prime}$ which are listed in Tab.3. The results are
about fifty times smaller than the experimental data.

In experimental fit, pole approximation is often used to fit the momentum
square dependence of form factor \cite{experiment}. As a try, we also assume
$F(k_1^2,k_2^2)=
\frac{1}{(1-k_1^2/m_{\eta^{\prime}}^2)(1-k_2^2/m_{\eta^{\prime}}^2)}$
for our calculations, we list the results in Tab.4. 
And we can see that this kind of form factor would make the digluon
fusion mechanism completely negligible in $B \rightarrow \eta^{\prime} K$.

To examine the validity of perturbative QCD in the above caculations, or in
other words, whether the amplitude calculated by perturbative QCD is dominated
by hard gluon contributions, we set cut on $k_2^2$ and compare the results
with those without cut. Taking $\Lambda_{QCD} \simeq 200 MeV$, we list the
numerical results in Tab.V. We can see that hard gluon contributions are
really dominant in the case of constant form factor. But if we take form
factor as $F(k_1^2, k_2^2)=\frac{m^2_{\eta^{\prime}}}{2 k_1 \cdot k_2}$, hard gluon
contributions are small, this 
 does not mean that non-perturbative contributions are dominant in this
case.
This is due to the fact that the singularity in the form factor
is accidently close to the other singularities in $k_1^2$ and $k_2^2$ and then 
enhance the contributions of soft gluon.

\section{Remarks and Conclusions}
The authors of Ref \cite{yangyd} proposed digluon fusion mechanism to explain
the large branching ratios of $B \rightarrow \eta^{\prime} K$, but because
of our ignorance about effective $\eta^{\prime}gg$ 
vertex, there are large uncertainties in perturbative QCD estimations.
We try several kinds of $\langle gg | \eta^{\prime} \rangle $ form factors, 
and through our calculations, we find that constant form factor is 
consistent with the data of $B \rightarrow \eta^{\prime} K$, but inconsistent 
with the data of $B \rightarrow \eta^{\prime} K^{\ast}$  in some
cases of different $\alpha_s$ or $\epsilon_B$ . if we take
$F(k_1^2, k_2^2)=\frac{m^2_{\eta^{\prime}}}{2 k_1 \cdot k_2}$ as the authors of
Ref \cite{yangyd} have done, di-gluon fusion mechanism is important in 
$B \rightarrow \eta^{\prime} K$ but not dominant. As a consequence, we
could anticipate that ${\cal{B}}(B \rightarrow \eta^{\prime} K^{\ast})$ would
be about $10^{-5}$ which can be tested by future experiments. We also try the 
other three kinds of form factors and they all give very small 
contributions to the branching ratios.

We conclude that, though we know little about 
$\langle gg | \eta^{\prime} \rangle$ form factor,
if di-gluon fusion mechanism is important in $B \rightarrow \eta^{\prime} K$,
the branching ratios of $B \rightarrow \eta^{\prime} K^{\ast}$ would be 
definitely around $10^{-5}$.

\section{Acknowledgements}
This work is supported in part by the National Natural Science Foundation of
China and the grant of China State Commission of Science and Technology.

\begin{table}
\vspace*{1cm}
\begin{tabular}{|c|c|c|c|c|c|c|c|c|c|c|}
$\epsilon_B$ & \multicolumn{3}{c|}{$0.05$} & \multicolumn{3}{c|}{$0.06$} 
& \multicolumn{3}{c|}{$0.07$} & Exp.\cite{CLEO} \\\cline{1-10}
$\alpha_s$ & $0.21$ & $0.28$ & $0.38$ & $0.21$ & $0.28$
& $0.38$ & $0.21$ & $0.28$ & $0.38$ & \\\cline{1-10}\cline{10-11}
${\cal{BR}}(B^{-}\rightarrow K^{-} \eta^{\prime})$ &
$1.99$ & $6.60$ & $21.3$ & $1.32$ & $4.40$ & $14.2$ & 
$0.94$ & $3.13$ & $10.1$ & $8.0^{+1.0}_{-0.9}\pm 0.8$ \\\hline
${\cal{BR}}({\bar{B}}^0 \rightarrow {\bar{K}}^0 \eta^{\prime})$  &
$2.06$ & $6.88$ & $22.2$ & $1.38$ & $4.59$ & $14.8$ & 
$0.98$ & $3.26$ & $10.5$ & $8.8^{+1.8}_{-1.6}\pm 0.9$ \\\hline
${\cal{BR}}(B^{-}\rightarrow K^{\ast -} \eta^{\prime})$ &
$4.04$ & $13.5$ & $43.4$ & $2.40$ & $8.00$ & $25.8$ &
$1.59$ & $5.30$ & $17.1$ & $<8.7$ \\\hline
${\cal{BR}}({\bar{B}}^0 \rightarrow {\bar{K}}^{\ast 0} \eta^{\prime})$ &
$4.19$ & $14.0$ & $45.1$ & $2.49$ & $8.31$ & $26.8$ &
$1.66$ & $5.52$ & $17.8$ & $<2.0$ 
\end{tabular}
\vspace*{0.5cm}
\caption{The branching ratios of $B\rightarrow K \eta^{\prime}$
and $B\rightarrow K^{\ast} \eta^{\prime}$ in unit of $10^{-5}$ by
using constant form factor $F(k_1^2,k_2^2)=1$. Where 
$\alpha_s(k_1^2 =12 GeV^2)=0.21$, $\alpha_s(k_2^2=1 GeV^2)=0.38$ and 
$\alpha_s=\sqrt{\alpha_s(k_1^2)\alpha_s(k_2^2)}=0.28$.} 
\end{table}

\begin{table}
 \vspace*{1cm}
\begin{tabular}{|c|c|c|c|c|c|c|c|c|c|c|}
$\epsilon_B$ & \multicolumn{3}{c|}{$0.05$} & \multicolumn{3}{c|}{$0.06$}
& \multicolumn{3}{c|}{$0.07$} & Exp. \cite{CLEO} \\\cline{1-10}
$\alpha_s$ & $0.21$ & $0.28$ & $0.38$ & $0.21$ & $0.28$
& $0.38$ & $0.21$ & $0.28$ & $0.38$ & \\\cline{1-10}\cline{10-11}
${\cal{BR}}(B^{-}\rightarrow K^{-} \eta^{\prime})$ & 
$0.60$ & $2.07$ & $6.42$ & $0.47$ & $1.63$ & $5.05$ &
$0.39$ & $1.34$ & $4.15$ & $8.0^{+1.0}_{-0.9}\pm 0.8$ \\\hline
${\cal{BR}}({\bar{B}}^0 \rightarrow {\bar{K}}^0 \eta^{\prime})$  &
$0.62$ & $2.15$ & $6.67$ & $0.49$ & $1.70$ & $5.27$ &
$0.40$ & $1.39$ & $4.31$ & $8.8^{+1.8}_{-1.6}\pm 0.9$ \\\hline
${\cal{BR}}(B^{-}\rightarrow K^{\ast -} \eta^{\prime})$ &
$0.91$ & $3.17$ & $9.83$ & $0.67$ & $2.31$ & $7.16$ &
$0.54$ & $1.89$ & $5.86$ & $<8.7$ \\\hline
${\cal{BR}}({\bar{B}}^0 \rightarrow {\bar{K}}^{\ast 0} \eta^{\prime})$ &
$0.95$ & $3.30$ & $10.2$ & $0.69$ & $2.40$ & $7.44$ &
$0.57$ & $1.97$ & $6.11$ & $<2.0$
\end{tabular}
\vspace*{0.5cm}
\caption{The branching ratios of $B\rightarrow K \eta^{\prime}$ and
$B\rightarrow K^{\ast} \eta^{\prime}$ in unit of $10^{-5}$ by using 
the form factor $F(k_1^2,k_2^2)=\frac{m_{\eta^{\prime}}^2}{2(k_1\cdot 
k_2)}$. Where
$\alpha_s(k_1^2=12 GeV^2)=0.21$, $\alpha_s(k_2^2=1 GeV^2)=0.38$ and   
$\alpha_s=\sqrt{\alpha_s(k_1^2)\alpha_s(k_2^2)}=0.28$.} \end{table}

\begin{table}
\vspace*{1cm}
\begin{tabular}{c|ccccc}
$\epsilon_B$ & $0.04$ & $0.05$ & $0.06$ & $0.07$ & $0.08$ \\\hline
${\cal{B}}(B^{-}\rightarrow K^{-} \eta^{\prime})$ &
$2.31$ & $1.71$ & $1.34$ & $1.09$ & $0.92$ 
\end{tabular}
\vspace*{0.5cm}
\caption{The branching ratios of $B\rightarrow K \eta^{\prime}$
in unit of $10^{-6}$ by using
the form factor $F(k_1^2,k_2^2)=(1-X^2/\Lambda^2)^{-1}$.}
\end{table}

\begin{table}
\vspace*{1cm}
\begin{tabular}{c|ccccc}
$\epsilon_B$ & $0.04$ & $0.05$ & $0.06$ & $0.07$ & $0.08$ \\\hline
${\cal{BR}}(B^{-}\rightarrow K^{-} \eta^{\prime})$ &
$2.01$ & $1.30$ & $0.91$ & $0.67$ & $0.52$ \\\hline
${\cal{BR}}({\bar{B}}^0 \rightarrow {\bar{K}}^0 \eta^{\prime})$ &
$2.09$ & $1.35$ & $0.95$ & $0.70$ & $0.53$ 
\end{tabular}
\vspace*{0.5cm}
\caption{The branching ratios of $B\rightarrow K \eta^{\prime}$
in unit of $10^{-6}$ by using the form factor
$F(k_1^2,k_2^2)=\frac{1}{(1-k_1^2/m_{\eta^{`}})(1-k_2^2/m_{\eta^{\prime}})}$.}
\end{table}

\begin{table}
\vspace*{1cm}
\begin{tabular}{c|c|c|c|c}
$Decay~Modes$ & \multicolumn{2}{c}{$B^{-}\rightarrow \eta^{\prime} K^{-}$} &
\multicolumn{2}{c}{${\bar{B}}^{0}\rightarrow \eta^{\prime} K^{\ast -}$} 
\\\hline
$Form ~Factor$
& $F(k_1^2,k_2^2)=\frac{m_{\eta^{\prime}}^2}{2(k_1 \cdot k_2)}$ &
$F(k_1^2,k_2^2)=1$ &
$F(k_1^2,k_2^2)=\frac{m_{\eta^{\prime}}^2}{2(k_1 \cdot k_2)}$ &
$F(k_1^2,k_2^2)=1$ \\\hline
$\Lambda_{QCD}^2$ & $71\%$ & $100\%$ & $72\%$ & $94\%$ \\\hline
$4\Lambda_{QCD}^2$ & $45\%$ & $93\%$ & $43\%$ & $84\%$ \\\hline
$9\Lambda_{QCD}^2$ & $22\%$ & $71\%$ & $17\%$ & $58\%$ 
\end{tabular}
\vspace*{0.5cm}
\caption{When $\epsilon_B=0.06$, the ratio of `hard contricbution' to
di-gluon fusion mechanism with $k_2^2$ cut of $\Lambda_{QCD}^2$,
$4\Lambda_{QCD}^2$ and $9\Lambda_{QCD}^2$.}
\end{table}


\begin{figure}[tb]
\vspace*{2cm}
\hspace*{-6cm}
\centerline{\epsfig{figure=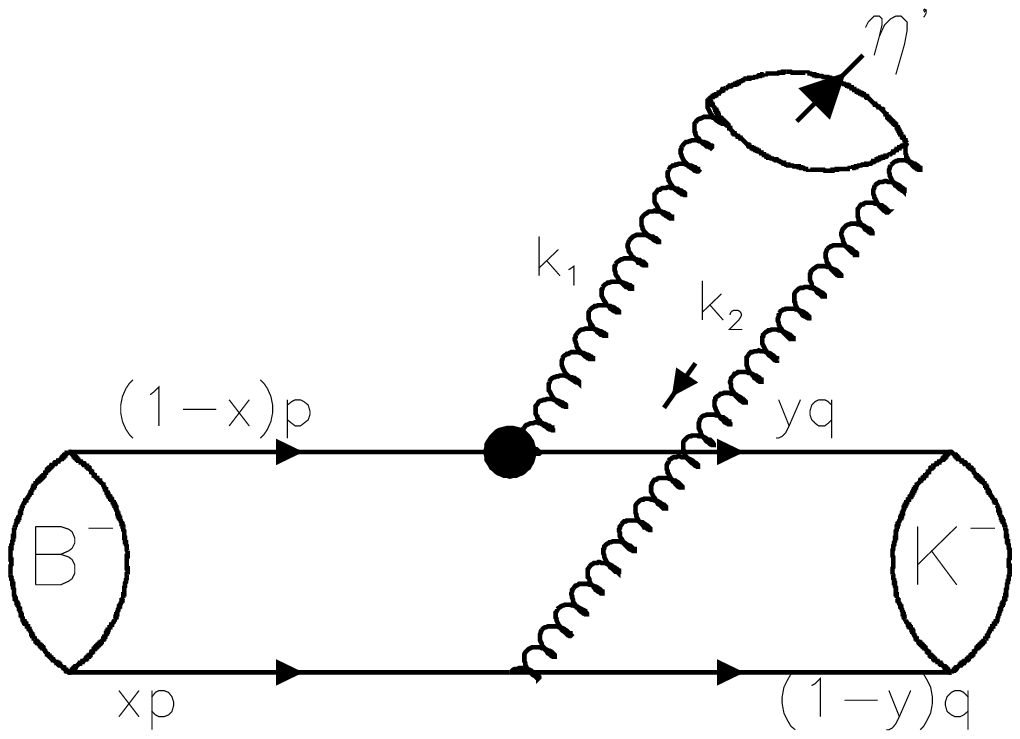,height=5cm,width=8cm,angle=0}}
\vspace*{-4cm}
\caption{\em The diagram for di-gluon fusion mechanism.}
\end{figure}


\begin{thebibliography}{17}
\bibitem{CLEO}
CLEO Collaboration, S.J. Richichi et al., CLEO-CONF-99-12, Aug 1999.
(preprint hep-ex/9908019).

\bibitem{Zhitnisky}
I. Halperin and A. Zhitnisky, Phys. Rev. Lett.{\bf{80}}, 438(1998).

\bibitem{He}
A. Datta, X.G. He and S. Pakvasa, Phys. Lett. B {\bf{419}}, 369(1998).

\bibitem{AS}
D. Atwood and A. Soni, Phys.Lett.B {\bf{405}}, 150(1997).

\bibitem{HT}
W. S. Hou and B. Tseng, Phys.Rev.Lett. {\bf{80}}, 434(1998).

\bibitem{CLEO1}
CLEO Collaboration, T.E. Browder {\it {et al}}., 
Phys. Rev. Lett.{\bf{81}}, 1786(1998).

\bibitem{yangyd}
D. Du, C. S. Kim and Y. Yang, Phys.Lett.B {\bf{426}}, 133(1998).


\bibitem{Hou}
W. S. Hou, Nucl. Phys. B {\bf{308}}, 561(1988).

\bibitem{Close}
F. E. Close, G. R. Farrar and Z. Li, Phys.Rev.D {\bf{55}}, 5749(1997).

\bibitem{wf}
A. Szczepaniak, E. Henley and S. J. Brodsky, Phy.Lett.B {\bf{152}}, 380(1990);\\
C. E. Carlson and J. Milana, Phys.Rev.D {\bf{49}}, 5908(1994); Phys.Lett.B
{\bf{301}}, 237(1993);\\
H. Simma and D. Wyler, Phys.Lett.B {\bf{272}}, 395(1991);\\
C. D. L\"u and D. X. Zhang, Phys.Lett.B {\bf{400}}, 188(1997).

\bibitem{kagan}
A. L. Kagan and A. A. Petrov, preprint hep-ph/9707354.

\bibitem{chenghy}
H.Y. Cheng and B. Tseng, Phys. Lett. B {\bf{415}}, 263(1997); 
H.Y. Cheng, hep-ph/9712244.

\bibitem{brodsky}
G. P. Lepage and S. J. Brodsky, Phys. Rev. Lett. {\bf{43}}, 545(1979);
Phys. Rev. D {\bf{22}}, 2157(1980); Phys. Rev. D {\bf{24}}, 1808(1981).

\bibitem{kuhn}
J. G. K\"{o}ner, J. H. K\"{u}hn, M. Krammer and H. Schneider,
Nucl. Phys. B {\bf{229}}, 115(1983).

\bibitem{experiment}
CLEO Collaboration, J. Gronberg {\it{et al}}., Phys.Rev.D {\bf{57}}, 33(1998);
L3 Collaboration, M. Acciarri {\it{et al}}., Phys.Lett.B {\bf{418}}, 399(1998).

\bibitem{ong}
Saro Ong, Phys.Rev.D {\bf{52}}, 3111(1995);
P. Kessler and S. Ong, Phys. Rev. D {\bf{48}}, 2974(1993).

\bibitem{He1}
Xiao-Gang He and Guey-Lin Lin, preprint hep-ph/9812248.
\end{thebibliography}
\end{document}